\documentclass[12pt]{article}
\usepackage[margin=1in]{geometry}
\usepackage{graphicx}
\usepackage{subcaption}
\graphicspath{{./plots/}}
\usepackage{float}
\usepackage{amsmath}
\usepackage[none]{hyphenat}
\usepackage[hidelinks,backref=page]{hyperref}
\renewcommand*{\backref}[1]{}
\renewcommand*{\backrefalt}[4]{%
	\ifcase #1 (Not cited.)%
	\or        (Cited on page~#2.)%
	\else      (Cited on pages~#2.)%
	\fi}
\usepackage[numbers]{natbib}
\usepackage{amssymb}
\usepackage{url} 

\usepackage{amsthm}

\newtheorem*{Thm*}{Theorem}

\usepackage[nottoc,notlot,notlof]{tocbibind}
\usepackage[title,titletoc]{appendix}

\usepackage{makecell}

\usepackage{enumitem}

\newcommand{\R}{\mathbb{R}}
\newcommand{\E}{\mathbb{E}}
\renewcommand{\P}{\mathbb{P}}

\DeclareMathOperator*{\argmax}{arg\,max}

\makeatletter
\g@addto@macro\bfseries{\boldmath}
\makeatother

\makeatletter
\newcommand*{\rom}[1]{\expandafter\@slowromancap\romannumeral #1@}
\makeatother

\usepackage[title]{appendix}

\makeatletter
\def\namedlabel#1#2{\begingroup
	\def\@currentlabel{#2}
	\label{#1}\endgroup
}
\makeatother

\begin{document}
	
	\title{The $\kappa$-generalised Distribution for  Stock Returns}
	\author{Samuel Forbes}
	\date{}
	
	\maketitle
	
	\begin{abstract}

			Empirical evidence shows stock returns are often heavy-tailed rather than normally distributed. The   $\kappa$-generalised distribution, originated in the context of statistical physics by  Kaniadakis, is characterised by the $\kappa$-exponential function that is asymptotically exponential for small values and asymptotically power law for large values. This proves to be a useful property and makes it a good candidate distribution for many types of quantities. In this paper we focus on fitting historic daily stock returns for the FTSE 100 and the top 100 Nasdaq stocks. Using a Monte-Carlo goodness of fit test there is evidence that the  $\kappa$-generalised distribution is a good fit for a significant proportion of the 200 stock returns analysed. 

	\end{abstract}
	
	\section{Introduction}
	
	Assuming stock returns are normally distributed and independent is the basis for their modelling with geometric Brownian motion as is the case in the Black-Scholes model \cite{black1973pricing}. However Mandelbrot \cite{Mandelbrot1963Variation} and Fama \cite{fama1965behavior} were amongst the first to use empirical evidence to show that stock returns are not in general normally distributed but are instead heavy-tailed. In particular they focused on fitting stocks with heavy-tailed stable distributions. However there has been debate on whether stock returns are stable \cite{lau1990distribution} and there seems to be no clear consensus on the distribution(s) that best describe stock returns. 
	
	The $\kappa$-generalised distribution discovered by Kaniadakis in a statistical physics context \cite{kaniadakis2001non} is a promising distribution for a number of regularly varying quantities. This may be due to the fact that the distribution is asymptotically exponential/Weibull for small values and  asymptotically power law for large values.
	We fit the negative and positive tails of the FTSE 100 and the top 100 Nasdaq historic daily stock returns with the $\kappa$-generalised distribution using numerical maximum likelihood estimation. Then using a Monte-Carlo goodness of fit test with the Kolmogorov-Smirnov statistic we find evidence that a significant proportion of the daily stock returns follow a $\kappa$-generalised distribution. We also fit with the stable distribution (of which the normal distribution is a particular case) and find these to be inferior at least with the parameterisation and methods we use here.

	\section{Stock Returns}
	
	\subsection{Definition}
	
	Let $X_t\in \R_{>0}$ be the price of a stock at time $t \in \R_{>0}\,.$ Define the \textbf{return of a stock}, as in \cite{fama1965behavior}, at time $t$ as
	\begin{equation}
		R_t = \log \left(\frac{X_t}{X_{t-\tau}} \right)\in \R
		\label{return}
	\end{equation}
	where $ \tau \in \R_{>0}$ is the time interval between two consecutive stock values. In this paper $\tau$ will be set to \textbf{one day}. Fitting to different time periods would be of interest for future study.
	
%
	
	To find the stock value from the return we can rearrange \eqref{return} to find 
	\begin{equation}
		X_t = X_{t-\tau}\exp(R_t)\,.
	\end{equation}
	We shall assume that the $R_t \sim D(\Theta)$ are identically distributed with distribution $D$ and parameters $\Theta\,$. If the $R_t \sim \mathcal{N}(\mu, \sigma^2)$ are normally distributed and independent then $X_t$ is lognormally distributed dependent on $X_{t-\tau}$ and can be modelled by geometric Brownian motion (GBM). However evidence suggests $R_t$ is instead in general heavy-tailed meaning GBM may not be such a good model for stock prices especially for  extreme jumps. 
	
	\subsection{The Tails of Stock Returns}
	
	We shall want to consider the two tails of stock returns:
	\begin{equation*}
		\begin{cases}
			\P(R_t<r) \,,& r<0 \,,\\
			\P(R_t>r) \,,& r>0 \,.
		\end{cases}
	\end{equation*}
	Stable distributions in general have support $\R$ so we can fit directly to the stock returns.
	However the $\kappa$-generalised distribution as defined in Section \ref{sec_k_gen} is supported
	on $(0,\infty)$ so we fit to the two tails separately. To do this
	we define random variables for the negative and positive returns as follows:
	\begin{equation*}
		R_t^{<} = \{R_t \,|\, R_t<0\}\,, \quad R_t^{>} = \{R_t \,|\, R_t>0\} \,.
	\end{equation*}
	Let $q = \P(R_t<0)$ then we can write the tails of stock returns as follows:
	\begin{align*}
		\P(R_t<r) &= q \P(R_t^{<} <r) = q \P(-R_t^{<} >-r)\,, \quad r<0\,, \\
		\P(R_t>r) &= (1-q) \P(R_t^{>} >r) \,, \quad r>0 \,.
	\end{align*}
	Thus for the $\kappa$-generalised distribution we estimate $q$ and then fit to $-R_t^{<}$ and 
	$R_t^{>} $ that both have support $(0,\infty)\,.$

	\section{Stable Distribution}
	
	A stable distribution $X$ is one such that adding independent copies of the distribution gives the same distribution. The stable distribution, apart from particular cases, has no analytical density or distribution function and is represented by the characteristic function. The distribution can be represented by a number of parameterisations, see Nolan \cite{nolan2020univariate}. We use the $0$\textsuperscript{th} parameterisation in  \cite{nolan2020univariate} denoted as
	$X \sim S(\alpha,\beta,\gamma,\delta;0)$ with four parameters 
	$\alpha \in (0,2]\,,\, \beta \in [-1,1]\,,\, \gamma \geq 0 \,,\, \delta \in \R\,.$ The characteristic function of this parameterisation is (see Definition 4 of \cite{nolan2020univariate})
	\begin{equation*}
		\phi_X(t) := \E[e^{itX}] 
		= \begin{cases}
			\exp(-\gamma^{\alpha} |t|^\alpha(1+i\beta \tan(\pi\alpha/2)\text{sign}(t)(|\gamma t|^{1-\alpha}-1))+i\delta t)\,,& \alpha \neq 1 \\
			\exp(-\gamma |t|(1+i \beta \dfrac{2}{\pi} \text{sign}(t) \log (\gamma|t|))+i\delta t)\,,& \alpha = 1
		\end{cases}
	\end{equation*}
	We note the following particular cases of the stable distribution \cite{nolan2020univariate}:
	\begin{enumerate}
		\item When $\alpha=2\,, \beta = 0$, $X$ is a normal distribution.
		\item When $\alpha=1\,, \beta = 0$, $X$ is a Cauchy distribution.
		\item When $\alpha=1/2\,, \beta = 1$, $X$ is a L\'{e}vy distribution.
	\end{enumerate}
	A property of particular interest is that the tails of $X$ are asymptotically power law under the condition that $\alpha<2\,$ (see Theorem 2.1 in \cite{nolan2020univariate}):
	\begin{equation} \label{ht_stable_tails}
		\P(X<-x) \underset{x \rightarrow \infty} {\sim} \gamma^{\alpha} c_{\alpha} (1-\beta) x^{-\alpha}\,,
		\quad \P(X>x) \underset{x \rightarrow \infty} {\sim} \gamma^{\alpha} c_{\alpha} (1+\beta) x^{-\alpha}
	\end{equation}
	where $c_{\alpha} = \sin(\pi\alpha/2)\Gamma(\alpha)/\pi$ and for general functions $f,\,g\,,$ $$f(x)\underset{x \rightarrow a} {\sim} g(x) \Leftrightarrow \lim_{x\rightarrow a}f(x)/g(x) =1\,.$$
	Thus we see by \eqref{ht_stable_tails} that the power law exponent $\alpha$ is the same in either tail. This and the fact that $\alpha<2$ are restrictions when fitting stock returns which are not found when fitting with the $\kappa$-generalised distribution separately to each tail. 
	
	\section{$\kappa$-generalised Distribution} \label{sec_k_gen}
	
	The $\kappa$-generalised distribution has been applied across many fields including high energy physics \cite{kaniadakis2005statistical}, the income and wealth distribution \cite{clementi2016kappa} and finance \cite{trivellato2013deformed}.
	To set up the distribution we define the  \textbf{$\kappa$-exponential function} as \cite{kaniadakis2002statistical}
	\begin{equation}
		\exp_{\kappa}(x) = (\sqrt{1+\kappa^2x^2}+\kappa x)^{1/\kappa}
		\label{k_exp}
	\end{equation}
	for some parameter $\kappa \in \R\,.$
	Discussion on the statistical physics origins and properties of \eqref{k_exp} are summarised in \cite{kaniadakis2002statistical}. In particular \eqref{k_exp} can be found by maximising the entropy of  it's inverse, the $\kappa$-logarithm $\log_{\kappa}(x) = (x^{\kappa}-x^{-\kappa})/(2\kappa)\,,$ under the restriction that $|\kappa|<1\,$, see Section \rom{5} of \cite{kaniadakis2002statistical}.
	Two notable features of the $\kappa$-exponential  function are that it asymptotically approaches a regular exponential function for small $x$ and  asymptotically approaches  a power law for large absolute $x$. Specifically
	
	\begin{equation}
		\exp_{\kappa}(x) \underset{x \rightarrow 0} {\sim}\exp(x) \, , \quad 
			\exp_{\kappa}(x) \underset{x \rightarrow \pm \infty} {\sim} |2\kappa x|^{\pm 1/|\kappa|} \, .
			\label{k_gen_ass}
	\end{equation}

	Now the particular version of the \textbf{$\kappa$-generalised distribution} we will use is used for income in \cite{clementi2007kappa}. Let $X$ be a continuous random variable on the support $(0,\infty)\,.$ Then we define the $\kappa$-generalised distribution by the tail 
	\begin{equation}
			\P(X>x) = \exp_{\kappa}(-\beta x^{\alpha}) \,, \quad x>0 \, , \label{k_gen_dist_tail}
	\end{equation}
	with parameters $\kappa \in (0,1),\,\alpha,\,\beta>0\,.$ Then by properties \eqref{k_gen_ass} we have 
	\begin{equation*}
		P(X>x)  \underset{x \rightarrow 0^{+}} {\sim} \exp(-\beta x^{\alpha})\, , \quad 
		P(X>x) 	\underset{x \rightarrow \infty} {\sim} (2 \kappa \beta)^{-1/\kappa}x^{-\alpha/\kappa}\, .
	\end{equation*}
	
	Thus the $\kappa$-generalised distribution is approximated by a Weibull distribution for small $x$ and is asymptotically power law for large $x$. Unlike for the stable distribution we have no upper bound on the power law exponent $\alpha/\kappa\,.$ The density of the 
	$\kappa$-generalised distribution can be found (taking the negative of the derivative of \eqref{k_gen_dist_tail}) to be
	\begin{equation}
		f(x; \alpha, \beta) = \frac{\alpha \beta x^{\alpha-1}  \exp_{\kappa}(-\beta x^{\alpha})}{\sqrt{1+\beta^2 \kappa^2 x^{2 \alpha}}} \,. \label{k_gen_dens}
	\end{equation}
	
	\section{Fitting}
	
	\subsection{Data} \label{sec_data}
	
 We have days $i=1, 2, \dots, n$ where day $n$ is the closest to the present day and each day $n-j$ is 
 $j = 1,2, \dots, n-1$ days prior to day $n\,.$ Let $S_i$ denote the price of the stock on day $i=1,2,\dots,n\,,$ then the $n$ stocks can be mapped to the $n-1$ returns $R_i$ , $i=2,3,\dots,n$ as follows
	\begin{equation*}
		(S_1, S_2, \dots, S_n) \rightarrow (R_2, R_3, \dots, R_n)
	\end{equation*}
	where 
	\begin{equation*}
		R_i = \log(S_i/S_{i-1}) \,.
	\end{equation*}
	
	The stock data we use is from the FTSE 100 and top 100 Nasdaq. We take day $n$ to be the 14\textsuperscript{th} May 2024 and day $1$ to be the first day we have a record of the stock price. We take the stock values as the closed value for each day. We obtain the stock data using the Python package yfinance \cite{y_finance} that extracts stock data from Yahoo finance. This data overall seems to be accurate however for one stock the author noticed some unusual fluctuations so we caution that there may be some inaccuracies in the data.
	
	\subsection{MLE and Goodness of Fit}
	
	\subsubsection{MLE}  \label{sec_MLE}
	
	We fit the returns data 
	for each stock using maximum likelihood estimation (MLE) using the normal distribution, the heavy-tailed stable distribution and the $\kappa$-generalised distribution. For the normal and stable distributions there are inbuilt Python SciPy functions to fit the distributions with MLE. For the $\kappa$-generalised distribution we write our own using a Python SciPy numerical solver to maximise the likelihood using the density \eqref{k_gen_dens}. In general MLE involves fitting the parameters of a chosen distribution by obtaining the maximum of the log-likelihood. Thus
	\begin{equation}
		\hat{\Theta} = \argmax_{\Theta} \left(  \sum_{i} \log (f(x_i; \Theta)) \right) \label{MLE}
	\end{equation}
	where $f$  is the density of the assumed distribution, $\{x_i\}$ is the  data, $\Theta$ are the unknown parameters to be estimated
	and $\hat{\Theta} $ is the MLE parameter estimation. Sometimes \eqref{MLE} has an analytical solution as is the case for the normal distribution however for the stable distribution in general and the $\kappa$-generalised distribution numerical methods are needed.
	
	\subsubsection{Goodness of Fit} \label{sec_goodness_fit}
	
	We shall describe a general Monte-Carlo goodness of fit test described in Section 4.1 of Clauset et al. for the power law \cite{clauset2009power} and originally used for the normal distribution by Lilliefors \cite{lilliefors1967kolmogorov}.  Let the  chosen distribution be $X \sim D(\Theta)$ where $D$ denotes the distribution with parameters $\Theta\,.$ Let us denote the data we want to perform the test on as $\{x_1, x_2,\dots,x_n\}\,.$ Define the Kolmogorov-Smirnov (KS) statistic
	\begin{equation}
		S = \max_{x_i} |\hat{\P}(X>x_i)-\P(X>x_i)| \,\label{KS_stat} 
	\end{equation}
	where $\hat{\P}$ is the empirical tail or one minus the empirical cumulative distribution function. Then the goodness of fit test is as follows:
	
	\begin{enumerate}
		\item We first fit the parameters $\hat{\Theta}$ of the distribution $D$ to the data and calculate $S$
		\eqref{KS_stat}.
		\item Then we randomly generate $N$ data sets of size $n$ from the distribution $D(\hat{\Theta})\,.$
		\item For each of the generated data sets $i=1,2\dots,N$ we fit parameters $\hat{\Theta}_i$  of the distribution $D$ and calculate \eqref{KS_stat} denoted by $S_i\,.$
		\item We estimate the $p$-value for the probability that the KS-statistic is smaller for the data than for the simulated data:
		\begin{equation*}
			\P(S_i>S) \approx p=\sum_{i=1}^{N} \mathbf{1}_{S_i>S} \,
		\end{equation*}
		where $\mathbf{1}$ is the indicator function.
		\item If $p < \alpha \in (0,1)$ we reject the hypothesis at level $\alpha$ that the data is from the distribution $D\,.$ Equivalently, higher values of $p \geq \alpha$ means we accept the hypothesis and so the data is more likely from the distribution $D\,.$
	\end{enumerate}

	\subsection{Results}
	
	Code for this section is found on the author's GitHub \cite{forbes_code}.
	We fit the FTSE 100 and Nasdaq top 100 daily stock returns, see Section \ref{sec_data}, to the normal distribution, heavy-tailed stable distribution and $\kappa$-generalised distribution using MLE.
	 For the $\kappa$-generalised distribution we fit separately to the negative and positive tails. We compare the fits using the KS-statistic and find that in all 200 cases the $\kappa$-generalised distribution is a better fit than either the heavy-tailed stable or the normal distribution. We also find in over 95\% of cases the heavy-tailed stable distribution is a better fit than the normal distribution. A fairer comparison my be the truncated stable distribution  \cite{levy1998wealthy}  however in this case the power law exponent will still be less than two. We found evidence from fitting with the $\kappa$-generalised that this is not always the case.
	
\begin{figure}[H]
  		\centering
  		\captionsetup{justification=centering}
  		\includegraphics[width=\textwidth]{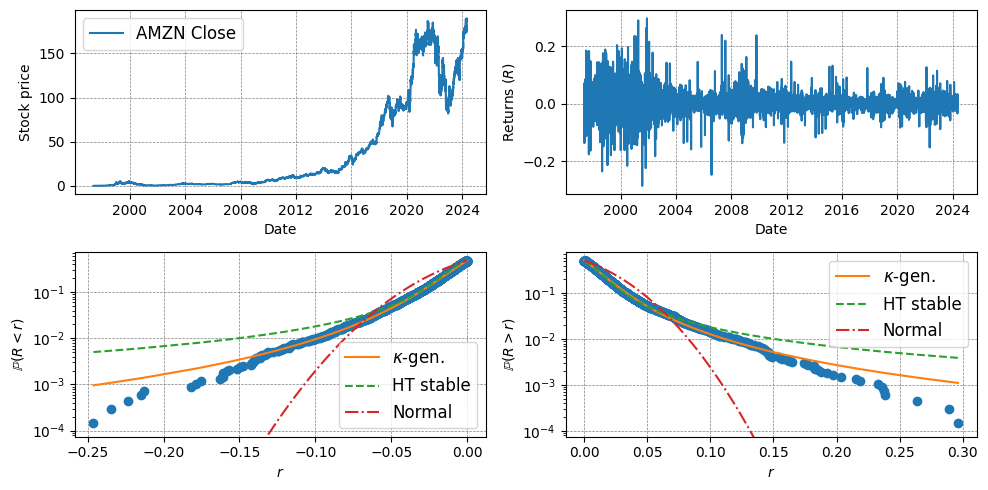}
  		\caption{Amazon daily historic stock prices (in dollars), returns \eqref{return} and normal, heavy-tailed stable and $\kappa$-generalised fits to the tails of the returns with logged y-scale. The $\kappa$-generalised fit is significant in both tails with the Monte-Carlo KS goodness of fit test.}
  		\label{AMZN}
  	\end{figure}

	We then use the Monte-Carlo goodness of fit test to compare with $N=100$ randomly generated data sets with the $\kappa$-generalised distribution to find how many of the 200 stocks are statistically significant with a $p$-value $p \geq \alpha=0.1\,.$ Due to the runtime of our code our value of $N$ is slightly low but we still feel it gives a reasonable indication of the significance levels. The conservative $\alpha = 0.1$ is also chosen in Clauset et al. \cite{clauset2009power}. Table \ref{table_gf} breaks down the percentage of stocks from the two indexes that are statistically significant with $\alpha=0.1$ to follow the $\kappa$-generalised distribution for the negative and positive returns. In particular we see that there is statistical evidence that roughly 39\% and 69\% of the stock returns from the FTSE 100 and Nasdaq top 100 can be fitted by the $\kappa$-generalised distribution for at least one of the tails. These stock codes are shown in Figure \ref{stock_codes}. We see that the $\kappa$-generalised distribution  works best for the Nasdaq top 100 daily stock returns. Figure \ref{AMZN} shows the fits to Amazon historic daily stock returns where we see visually that the $\kappa$-generalised distribution provides a good fit.
	
	\begin{table}[H] 
	\begin{center}
	\def\arraystretch{1.5}
	\begin{tabular}{ l | c | c}
		 & FTSE 100 & Nasdaq top 100 \\ \hline
		Negative returns & 24\% &  54\% \\
		Positive returns  &  33\% & 58\% \\
		Both negative and positive returns & 18\% &  43\% \\
		Either negative or positive returns  &  39\% &  69\% \\
	\end{tabular}
  	\end{center}
  	\caption{Rough percentage of stock returns that are statistically significant for following the  $\kappa$-generalised distribution with the Monte-Carlo KS test with $N=100$ and $\alpha=0.1\,$. This is broken down by the two indexes and by negative and positive returns. }
  		\label{table_gf}
  	\end{table}
  	
  	\begin{figure}[H]
  		\centering
  		\captionsetup{justification=centering}
  	\begin{verbatim}
  	 Significant FTSE 100 stocks either tail:
  	[`AAF' `AAL' `ADM' `AUTO' `AZN' `BATS' `BME' `BP.' `BRBY' `BT.A' `CCH'
  	`CNA' `CTEC' `DCC' `EXPN' `EZJ' `FLTR' `FRAS' `FRES' `GLEN' `GSK' `HIK'
  	`HLN' `IAG' `IHG' `IMB' `ITRK' `LLOY' `LSEG' `MNDI' `MNG' `NXT' `OCDO'
  	`PHNX' `RMV' `SHEL' `SKG' `STJ' `TSCO']
  		
	Significant Nasdaq top 100 stocks either tail:
	[`ABNB' `ADI' `ADSK' `AMGN' `AMZN' `ANSS' `ASML' `AVGO' `AZN' `BIIB'
	`BKNG' `BKR' `CDW' `CEG' `CHTR' `CPRT' `CRWD' `CSCO' `CSGP' `CTSH' `DASH'
	`DDOG' `DLTR' `DXCM' `FANG' `FAST' `FTNT' `GEHC' `GFS' `GILD' `GOOG'
	`GOOGL' `DXX' `ILMN' `INTU' `ISRG' `KDP' `KHC' `LIN' `LULU' `MAR' `MCHP'
	`MDB' `MDLZ' `MELI' `META' `MRNA' `MRVL' `MSFT' `NFLX' `NVDA' `ON' `ORLY'
	`PANW' `PAYX' `PDD' `PYPL' `QCOM' `ROP' `ROST' `SBUX' `TEAM' `TMUS'
	`TSLA' `TTD' `TTWO' `VRTX' `WBD' `WDAY']
  	\end{verbatim}
  		\caption{Stock codes of FTSE 100 and Nasdaq top 100 that are significantly fitted using Monte-Carlo KS goodness of fit test with $\kappa$-generalised distribution in either tail.}
  	\label{stock_codes}
  	\end{figure}
  	
	\newpage
	
	\section{Conclusion}
	
	This paper investigated how well the $\kappa$-generalised distribution fitted historic daily stock returns data. For this purpose we used the FTSE 100 and Nasdaq top 100. We found that the $\kappa$-generalised distribution, when compared using the KS statistic, was a better fit than a stable distribution in all cases. A significant proportion of the stock returns, particularly for the Nasdaq stocks, satisfied the KS Monte-Carlo goodness of fit test with a conservative $p$-value for the $\kappa$-generalised distribution.
	
	\bibliographystyle{apalike} 
	\bibliography{mybib}

\end{document}